\documentstyle[aps,preprint,epsf]{revtex}

\newcommand{\beq}{\begin{equation}}
\newcommand{\eeq}{\end{equation}}
\newcommand{\bea}{\begin{eqnarray}}
\newcommand{\eea}{\end{eqnarray}}

\begin{document}
\draft

\title{Overcharging of a macroion by an oppositely charged polyelectrolyte}
\author{T. T. Nguyen \and B. I. Shklovskii}
\address{
  Department of Physics, University of Minnesota, 116 Church St. Southeast,
  Minneapolis, MN 55455, USA
}

\maketitle

\begin{abstract}
Complexation of a polyelectrolyte with an
oppositely charged spherical macroion is studied for
both salt free and salty solutions.
When a polyelectrolyte winds around the macroion, its
turns repel each other and form an almost equidistant solenoid.
It is shown that this repulsive correlations of turns
lead to the charge inversion: more polyelectrolyte winds
around the  macroion than it is necessary
to neutralize it. The charge inversion becomes stronger with
increasing concentration of salt and can exceed 100\%. Monte-Carlo 
simulation results agree with our analytical theory.



\end{abstract}

\newpage

\section{Introduction}
Electrostatic interactions play an important role in
aqueous solutions of biological and synthetic polyelectrolytes (PE).
They result in the aggregation and complexation of oppositely charged
macroions in solutions. For example, in the chromatin,
negative DNA winds around a
positive histone octamer to form a complex known as the nucleosome.
The nucleosome was found to have negative net charge $Q^*$
whose absolute value is as large as 15\%
of the bare positive charge of the protein, $Q$.
This counterintuitive phenomenon is called the charge inversion 
and can be characterized by the charge inversion ratio, $|Q^*|/Q$.
For PE-micelle systems, charge inversion has been predicted by
Monte-Carlo simulations~\cite{Linse} and
observed experimentally~\cite{Dubin}.

These and other examples have recently stimulated several
theoretical studies of charge inversion accompanying the
complexation of a flexible PE with a rigid spherical
or cylindrical macroion of opposite sign~
\cite{Pincus,Bruinsma,Joanny,Sens}
(for more extensive bibliography on this subject
 see Ref.~\onlinecite{Bruinsma}).
All these authors arrive at the charge inversion
 for such a complexation.
It was also shown that if the PE molecule is
not totally adsorbed at the surface,
its remaining part is repelled by the inverted
charge of the macroion and
forms an almost straight radial
tail~\cite{Pincus,Bruinsma}(see Fig. 1).
However, all these papers use different models and seemingly
deal with charge inversion of different nature.
Surprisingly, both Refs.~\onlinecite{Pincus,Bruinsma} show that
the inverted charge of a macroion $Q^*$
does not depend on the value of the bare charge $Q$.

In this paper we present a new theory of complexation
of a flexible PE with an oppositely charged
rigid sphere. 
We consider here only the case of a weakly charged PE which does not
create Onsager-Manning condensation.
We show that both in
salt free and salty solutions the
charge inversion by such PE is driven by repulsive
correlations of PE turns at
the macroion surface. Such correlations
make an almost equidistant solenoid (see Fig. 1),
which locally resembles one-dimensional Wigner crystal along the
direction perpendicular to PE.
In the absence of salt, the charge inversion ratio
is smaller than 100\%. In a
salty solution, it grows with the salt concentration.
When the Debye-H\"{u}ckel screening radius
$r_s$ becomes smaller than the distance between
neighboring turns $A$,
the charge inversion ratio can be larger than 100\%.

The charge inversion of a macroion due to complexation with one PE
molecule can be explained in the way similar to 
Refs.~\onlinecite{Shklov99,Nguyen}, which dealt
with the charge inversion of a macroion screened
by many rigid multivalent counterions ($Z$-ions).
The tail repels adsorbed PE and creates correlation hole or,
in other words, its positively charged image. This image in the
already adsorbed layer of PE is responsible for the
additional correlation attraction to the surface,
which leads to the charge inversion. 

We show that smearing of charge PE on the surface 
of the sphere 
employed in Ref. \onlinecite{Pincus} is a good 
approximation only at $A \sim a$.
%
If $A \gg a$ smearing of charge at 
the surface of sphere is a rough approximation and leads to 
anomalously strong inversion
of charge and to the unphysical independence of the
inverted charge $Q^*$ on $Q$. The reason of
this phenomenon is easy to understand.
Smearing means that the PE solenoid is assumed 
to behave as a perfect metal. A neutral metal surface
can adsorb a charged PE due to image forces, making the
charge inversion ratio infinite. 
In reality, for an insulating macroion, an image of a
point charge in the PE coil can not be smaller than $A$ and the energy
of attraction to it vanishes at growing $A$. 
Only a macroion with a finite charge $Q$ adsorbs a PE coil with
a finite $A$. Therefore, $Q^*$
depends on $Q$ and the charge inversion ratio is always finite.

Our analytic theory is followed 
by Monte-Carlo simulations. They demonstrate good
agreement with the theory.
%

\section{An analytical theory.}

For a quantitative calculation, consider the complexation of a
negative PE with linear
charge density $-\eta$ and length $L$,
with a spherical macroion with radius $R$ and positive
charge $Q$. We assume that the PE is weakly charged,
i. e. $\eta \ll \eta_c$, where $\eta_c=k_BTD/e$ is Onsager-Manning
critical linear density, $T$ is the
temperature, $k_B$ is the Boltzmann constant and $D$ is the dielectric 
constant of water. In this case, 
there is no Onsager-Manning condensation of 
counterions and one can use linear theory of screening.
Because we are interested in the charge inversion
of the complex, we assume that
the PE length $L$ is greater
than the neutralizing length ${\cal L}=Q/\eta$.
In this case, a finite length $L_1$ of the PE
is tightly wound around the macroion due to the electrostatic
attraction.
The rest of the PE with length $L_2=L-L_1$
can be arranged into two possible configurations:
one tail with length $L_2$
or two tails with length $L_2/2$ going in opposite directions radially
outwards from the center of the macroion.
In both cases, the tails are straight to minimize
its electrostatic self-energy. We assume that ${\cal L} \gg R$, so that
there are many turns of the PE around the sphere.
Our goal is to calculate the net charge of the complex
$Q^*= Q - L_1\eta =({\cal L}-L_1)\eta$ 
and the charge inversion ratio $|Q^*|/Q$.
We show that, in the most common configuration with one tail,
this net charge
is negative: more PE winds around the macroion than
it is necessary to neutralize it.

Let us start from the salt free solution in which all
Coulomb interactions are not screened.
For simplicity, we assume that the PE has no
intrinsic rigidity, but its linear charge density
is large so that it has a rod-like configuration in solution
due to Coulomb repulsion between monomers.
When PE winds around the macroion, the strong
Coulomb repulsion between the neighboring  PE turns
keeps them parallel to each other and establishes an
almost constant distance $A$ between them (Fig. 1).
The total energy of the macroion with
the PE solenoid wound around it, $F_1$,
can be written as a sum of the Coulomb energy of
its net charge plus the self-energy of PE:
\beq
F_1 = (L_1-{\cal L})^2/2R + L_1\ln(A/a)~.
\label{energy}
\eeq
Here and below we write
all energies in units of $\eta^{2}/D$,
where $D$ is dielectric
constant of water (thus, all
energies have the dimensionality of length.)
The second term in Eq. (\ref{energy}) deserves 
special attention. The self-energy of a
straight PE of length $L_1$ in the solution is
$L_1\ln(L_1/a)$. However, when it winds around the macroion,
every turn is effectively screened by the neighboring turns
at the distance $A$.
This screening brings the self-energy down to $L_1\ln(A/a)$.
At length scale greater than $A$, the surface
charge density of the spherical complex is uniform
and the excess charge $L_1-{\cal L}$ is taken into
account by the first term in Eq. (\ref{energy}).
In other words, one can interpret Eq. (\ref{energy}) thinking
about our system as the superposition of a uniformly charged
sphere with charge $(L_1-{\cal L})$ and a neutral complex consisting
of the solenoid on a neutralizing spherical background. The total energy 
of these two objects is additive. Indeed, the energy of interaction 
between them vanishes because the first one creates a constant potential
on the second neutral one.

One can also rewrite the energy of solenoid on
 the neutralizing background as
\beq
L_1\ln(A/a) = L_1\ln(R/a) - L_1\ln(R/A)~.
\label{selfenergy}
\eeq
Here the first term is the self-energy of the PE with length $L_1$
whose turns are randomly positioned on the macroion.
(Indeed, for a strongly charged PE, each PE
turn is straight up to a distance of the order
of $R$ due to its electrostatic rigidity.
If we keep a PE turn fixed and average
over random positions of all other turns
we find our turn
on the uniform spherical background of opposite charge.
The absolute value of the background charge is of the order $R$,
the energy of interaction of our turn with it is of the order $R$
and is negligible compared to the turn's  self-energy $ R \ln(R/a)$
or $\ln(R/a)$ per unit length.)
Now it is easy to identify the second term of Eq. (\ref{selfenergy})
as the correlation energy. It represents
the lowering of the system's energy by forming an equidistant
coil from the random one. This correlation energy, $E_{cor}$, is of the order of
the interaction of the PE turn with its background (a stripe of
of the length $R$ and the width $A$
of the surface charge of the macroion) because all other turns
lie at the distance $A$ and beyond.
Estimating $A \sim R^2/L_1$,
we can write
\beq
E_{cor} \simeq -L_1\ln(R/A) \simeq -L_1\ln(L_1/R).
\label{Ecor}
\eeq
Substituting Eqs. (\ref{Ecor}) and (\ref{selfenergy}) into Eq.
(\ref{energy})
 for the total energy of the spherical complex, we obtain
\beq
F_1=L_1\ln(R/a)-L_1\ln(L_1/R)+(L_1-{\cal L})^2/2R~.
     \label{freecomplex}
\eeq
To take into account the PE tails,
let us consider each tail configuration separately.

{\it One tail configuration}. In this case,
the total free energy of the system is the sum of that
of the spherical complex, the self-energy of the tail and their
interaction. This gives:
\beq
F=F_1 + L_2 \ln(L_2/a)+ (L_1-{\cal L})\ln\left[(L_2+R)/R\right]~.
\label{freeone}
\eeq
To find the optimum value of the length $L_1$
one has to minimize $F$
with respect to $L_1$. Using Eq. (\ref{freeone}) and
the relation $L_2=L-L_1$, we obtain
\beq
(L_1-{\cal L})\left[R^{-1}-(L-L_1+R)^{-1}\right]=\ln({\cal L}/R)~,
\label{onetail}
\eeq
where we neglected terms of the order of unity and took into account
that $L_2 \gg R$ (as shown below, Eq. (\ref{l2c})).
The physical meaning of Eq. (\ref{onetail}) is transparent:
The left side
is the energy of the Coulomb repulsion of the net charge
of the spherical complex which has to be overcome in order to bring
an unit length of the PE from the tail to the sphere.
The right hand side (in which, $L_1$ has
been approximated by ${\cal L}$)
is the absolute value of the correlation energy
gained at the sphere which helps
to overcome this repulsion (See Eq. (\ref{Ecor})).
Equilibrium is reached when these two
forces 
are equal. From  Eq. (\ref{onetail}),
one can easily see that $L_1-{\cal L}$ is positive,
indicating a charge inversion scenario: 
more PE collapses on the macroion
than it is necessary to neutralize it.
Eq. (\ref{onetail}) also
clearly shows that correlations are the driving force of
charge inversion.

To understand how the length $L_1$ varies for different PE
length $L$, it is instructive to solve Eq. (\ref{onetail}) graphically.
One can see the following behavior (Fig. 2):

(a) When $L-{\cal L}$ is small,
Eq. (\ref{onetail}) has no solutions,  $\partial F/\partial L_1$
is always negative. The free energy is a monotonically decreasing
function
of $L_1$ and is minimal when $L_1=L$. In this
regime, the whole PE collapses on the macroion.

(b) As $L$ increases beyond a length $L^*$,
Eq. (\ref{onetail}) acquires two solutions,
which correspond to
a local minimum and a local maximum in
the free energy as a function of $L_1$.
The global minimum is still at $L_1=L$ and the whole PE remains
in the collapsed state.

(c) When $L$ increases further, at a length
$L=L_c$, the local minimum in the free energy at $L_1 < L$
becomes smaller
than the minimum at $L_1 = L$.
A first order phase transition happens
and a tail with a finite length $L_2$ appears.
$L_c$ can be found from the requirement that the
equation $F(L_1)-F(L)=0$ has
solutions at $0<L_1<L$. Using Eq. (\ref{freeone}), one gets
\beq
L_c\simeq {\cal L}+R\ln({\cal L}/R)+
R\sqrt{\ln({\cal L}/R)}\sqrt{\ln\ln({\cal L}/R)},
\label{lc}
\eeq
and the tail length $L_2$ at this critical point is
\beq
L_{2,c} \simeq R\sqrt{\ln({\cal L}/R)}\sqrt{\ln\ln({\cal L}/R)}~.
\label{l2c}
\eeq
As $L$ continues to increase, $L_1$ decreases and eventually
saturates
at the constant value
\beq
L_{1,\infty}={\cal L}+R\ln({\cal L}/R) \simeq L_c-L_{2,c}~,
\label{l1one}
\eeq
which can be found from Eq. (\ref{onetail}) by letting
$L\rightarrow\infty$.
Eq. (\ref{lc}) - (\ref{l1one}) are 
asymptotic results valid at ${\cal L/}/R \rightarrow\infty$.
If ${\cal L/}/R$ is not very large one can find $L_{1}(L)$
minimizing Eq. (\ref{freeone})
numerically. In Fig. 3 we present results for the case
${\cal L}=25 R$, which corresponds to $25/2\pi \simeq 4$
turns. In this case, $L_c=35.5 R$, $L_{2,c}=4.0 R$
and $L_{1,\infty}=30.4 R$. 
It should be also noted that, as Fig. 3 and
Eqs. (\ref{lc}), (\ref{l2c}) and (\ref{l1one}) suggest, 
$L_1$ is almost equal $L_{1,\infty}$ after the phase transition.

At ${\cal L} \gg R$, the charge inversion ratio 
$|Q^*|/Q = (L_1-{\cal L})/{\cal L}$ 
can be calculated from Eqs. (\ref{lc}) and (\ref{l1one}):
$|Q^*|/Q = (R/{\cal L})\ln({\cal L}/R) \ll 1$.
Thus, the charge inversion ratio is only logarithmically
larger than the inverse number of PE turns in the coil.

Using the insight gained above, we are now in a position
to achieve better understanding of the nature of the approximation
employed in Ref. \onlinecite{Pincus}.
The authors of Ref.~\onlinecite{Pincus} 
replaced the adsorbed PE by the same charge uniformly smeared 
at the macroion surface. Therefore, the term $L_1\ln(A/a)$ was omitted
in Eq. (\ref{energy}), so that at $A \gg a$, the correlation energy
was overestimated. This approximation replaces
the right hand side of Eq. (\ref{onetail}) by
the self-energy of a unit length of the tail. 
Correspondingly,
Eq. (\ref{onetail}) now balances the self-energy 
of a unit length of the tail with the electrostatic energy of 
this unit length smeared at the surface of overcharged macroion.
Thus we can call this mechanism of charge 
inversion ``the elimination of the self-energy" or simply ``metallization". 

As a result, the charge inversion obtained in Ref. 
\onlinecite{Pincus}, at $A\gg a$, is larger than that
of our paper. (Our correlation mechanism can be interpreted 
as a partial elimination of the self-energy. 
The second term of Eq. (\ref{energy})
is what is left from the PE self-energy due to self screening 
of PE at the distance $A$.)
Surprising independence of $Q^*$ on $Q$
or, in other words, the possibility of an infinite charge 
inversion ratio obtained in Ref.~\onlinecite{Pincus}
is also related to smearing of PE on the
macroion surface. This happens because when PE arrives at
the macroion surface it looses all its (positive) self-energy. 
This brings about an energy gain which does not depend 
on the bare charge of the macroion.

On the other hand, at $A\sim a$, the smearing of PE is a good approximation
and our results are close to that of Ref. \onlinecite{Pincus}.

{\it Two tails configuration}.
The free energy of the system can be written similar to
Eq. (\ref{freeone}), keeping in mind that we have two
tails instead of one, each with length $L_2/2$:
\bea
F&=&F_1+L_2 \ln\frac{L_2}{2a}+2(L_1-{\cal L})
\ln\frac{L_2+2R}{2R} + \nonumber \\
&& +(L_2+2R)\ln\frac{L_2+2R}{2R}-(L_2+4R)\ln\frac{L_2+4R}{4R}.
\label{freetwo}
\eea
The last two terms describe the interaction between the tails.
The optimum length $L_1$ can be found from the condition of a minimum
in the free energy.
Taking into account that, as shown below, $L_2 \gg R$ and ignoring
terms of the order unity, one gets
\beq
(L_1-{\cal L})
        \left[R^{-1}-(L_2/2+R)^{-1}\right]+
 \ln(L_2/R)=\ln({\cal L}/R)~.
\label{twotail}
\eeq
Comparing this equation to Eq. (\ref{onetail}), one finds an
additional potential energy cost $\ln(L_2/R)$ for bringing a unit
length
of the PE from the end of a tail to the sphere.
It originates from the interaction
of this segment with the other tail.
When $L$ is not very large, 
$L_2 \ll \cal L$,
one can neglect this
additional term and the two tail system behaves like
the one tail one. At a small
$L$, the whole PE lies on the macroion surface and the system is
overcharged. As $L$ increases,
eventually a first order phase transition happens, where two tails
with length of the
order $R\sqrt{\ln({\cal L}/R)}$ appear.
On the other hand, when $L$ is very large, such that 
$L_2 \gg \cal L$,
the new term dominates and the macroion becomes undercharged
($L_1-{\cal L}$ is negative)
with $L_1$ decreasing
as a logarithmic function of the PE length:
$L_1 \simeq {\cal L} -R\ln(L/{\cal L})$.
At an exponentially large value of
$L\sim {\cal L}\exp({\cal L}/R)$, the length $L_1$ reaches zero
and the whole PE unwinds from the macroion.

Above, we have described configurations with one
tail and two tails separately. One should ask which of them
is realized at a given $L$. Numerical calculations show
that, when $L$ is not very large,
the overcharged, one tail configuration is lower in energy.
At a very large value
of $L$, the complex undergoes a first order phase transition to
a two tails configuration and becomes undercharged. The value
of this critical length $L_{cc}$ can be estimated by
equating the free energies (\ref{freeone}) and (\ref{freetwo})
at their optimal values of $L_1$ 
which are ${\cal L}+R\ln({\cal L}/R)$ and ${\cal L}-R\ln(L/{\cal L})$ 
respectively.
In the limit where $\ln({\cal L}/R) \gg 1$, keeping only highest order 
terms, we get $L_{cc} \sim {\cal L}^2/R $, which indeed is
a very large length scale. 
This order of appearance of one and two tail configurations
is in disagreement with Ref.~\onlinecite{Pincus}.

In practical situations, there is always
a finite salt concentration in
a water solution. One, therefore, has to take the finite
screening length $r_s$ into account. 
For any reasonable $r_s$, $L_{cc} \gg r_s$,
and all Coulomb interactions responsible
for the transition from one to two tails
are screened out. Therefore, in a salty solution
the two tail configuration disappears.
Below we concentrate on the effect of screening
on one tail or tail-less configurations only.

In a weak screening case, when $r_s \gg L_{2,c}$,
Coulomb interactions responsible for the appearance
of the  tail remain unscreened.
Therefore, the lengths $L_c$ and
$L_{2,c}$ remain  almost unchanged. The
large $L$ limit of $L_1$ however should be modified. At a very large
tail length $L_2$ one should replace $L-L_1=L_2$ by $r_s$ in
 Eq.~(\ref{onetail}) because the potential vanishes beyond the distance
$r_s$. This gives
$$L_{1,\infty}(r_s)={\cal L}+R\ln({\cal L}/R)+
(R^2/r_s)\ln({\cal L}/R)~~.$$
One can see that $L_{1,\infty}$ increases
and charge inversion is stronger as $r_s$ decreases.
This is because when $r_s$ decreases,
the capacitance of the spherical complex
increases, the self-energy of it decreases
and it is easier to charge it.

When $R < r_s < L_{2,c}$, it is easy to show that
the tail length, which appears at the
phase transition, is equal to $r_s$ instead of $L_{2,c}$.
This means that, before a tail is driven out
at the phase transition, more PE condenses on
the macroion in a 
salty solution than that for the salt free case. In other
words, the critical point $L_c$ is shifted towards larger values:
$$L_c(r_s) = L_{1,\infty}(r_s) + r_s~~.$$
Obviously, $L_c(r_s) > L_c$ for $r_s < L_{2,c}$ and $L_c(r_s)$
approaches $L_c$ at $r_s\sim L_{2,c}$.
When $r_s$ approaches $R$, the critical length $L_c(r_s)$ reaches
${\cal L}+2R\ln({\cal L}/R)$, so that the inverted charge
is twice as large as that for the unscreened case.

At stronger screening, when $r_s < R$, to a first approximation,
the macroion surface can be considered as a charged plane.
The problem of adsorption of many rigid PE molecules
 on an oppositely charged plane has been
studied in Ref. \onlinecite{Nguyen}, where the role of
Wigner crystal like correlations similar to that
shown in Fig. 1 was emphasized.
The large electrostatic rigidity of a strongly charged PE
makes this calculation applicable to our problem as well.
One can use results of Ref. \onlinecite{Nguyen}
in three different ranges of $r_s$:
$R > r_s > A$, $A >r_s > a$, $a > r_s$.
In all these ranges, the net charge $Q^*$ of the macroion is
proportional to $R^2$
instead of an almost linear dependence on $R$ in
a salt free solution. The tail is not important
 for the calculation of the charge inversion ratio
because it produces only a local effect
near the place where the tail stems from the macroion.
Inverted net charge $Q^*$ grows with decreasing $r_s$, so that
charge inversion ratio of the macroion reaches 100\% at $r_s \sim A$
and can become even larger at $r_s \ll A$. 
For $r_s \ll A$ , our results are in agreement 
with those of Ref. \onlinecite{Joanny,Rubin}.
One should be aware that $|Q^*|$ ceases to increase at very small
$r_s$. This is because at an extremely small
$r_s$ such that the interaction between the macroion and
one persistence length of the PE becomes less than
$k_BT$, the PE desorbs from the macroion and the macroion
becomes undercharged. Therefore, $|Q^*|$ should reach a maximum at
a very small $r_s$ and then decrease.

Finally, it should be noted that in the above discussion of 
the role of screening, we neglected
the possibility of the condensation of the PE's counterions
on the sphere with inverted charge. This is valid for a large enough
screening length because it is well known that in this case  condensation
does not occurs on a spherical macroion. Using $Q^* \sim R\ln({\cal L}/R)$
and the standard condition for the condensation on a charged sphere
\cite{Gueron}, 
it is not difficult to show that the sphere is screened linearly
if
$$r_s > R^{1-\eta/2\eta_c}{\cal L}^{\eta/2\eta_c}~~.$$
When $r_s <R$, the macroion can be approximated as a charged plane
and it is also known that a planar charge is linearly screened
if the screening radius is small enough. Specifically, Eq. (73) of Ref.
\onlinecite{Nguyen} shows that screening is linear if
$$r_s < Ae^{\eta_c/\eta}\sim \frac{R^2}{\cal L}e^{\eta_c/\eta}~~.$$
As we can see, when $\eta$ is less than $\eta_c$ by a logarithmic factor,
i. e. when $\eta < \eta_c /\ln({\cal L}/R)$, the range of $r_s$,
where the macroion is nonlinear screened, almost vanishes. 
For $\eta$ of the order of $\eta_c$, however, there is a range
of $r_s$ where 
counterion condensation on the 
charge-inverted sphere 
has to be taken into account and the sphere's net
charge is different from our estimate. There are two aspects
of this counterion condensation phenomenon. Obviously, due to
stronger nonlinear screening at the sphere surface,
more PE collapses onto the sphere and the charge inversion ratio is
even larger
than what is predicted above in the linear screening theory.
On the other hand, if one defines the net charge of the sphere as
the sum of its bare charge, the charges of the collapsed PE monomers
and the charges of all counterions
condensed on it, the magnitude of this net charge is 
limited at the value given by the
theory of counterion condensation on a sphere~\cite{Gueron}.
As explained in Ref. \onlinecite{Nguyen}, it is this charge that 
is observed in electrophoresis.

Until now we talked about a weakly charged PE with $\eta \leq
\eta_c$.
In Ref. \onlinecite{Nguyen} we studied adsorption
of a strongly charged PE (for e.g., DNA) with $\eta \gg \eta_c$ on
positively charged plane.
Such PE initiates Onsager-Manning counterion condensation both in the
bulk and at the plane.
The theory Ref. \onlinecite{Nguyen} can be applied for the sphere at
$r_s \ll R$, too.
It predicts a strong charge inversion which grows with decreasing $r_s$
and exceeds 100\% at $r_s < A$.

\section{Monte-Carlo simulations.}

To verify the results of our analytical theory, we do Monte Carlo (MC) 
simulations. 
The PE is modeled as a chain of $N$ freely jointed
hard spherical beads each with charge $-e$ and radius
$a=0.2l_B$ where $l_B=7.12$\AA~is the Bjerrum length at room
temperature $T_{rm}=298^{\,\rm{o}}{\rm K}$ in water.
The bond length is kept fixed and equal
to $l_B$, so that our PE charge density $\eta$
is equal
to the Manning condensation critical charge density $\eta_c=k_BT_{rm}D/e$. 
Due to the discrete nature of the simulated PE, 
in order to compare simulation
results with theoretical predictions,
we refer to
the number of monomers $N$ as the PE length $L$ measured in units of $l_B$.
The macroion is modeled as a sphere
of radius $4l_B$ and with charge $100e$ uniformly distributed at 
its surface.
To arrange the configuration of the PE globally,
the pivot algorithm is used. In this algorithm, a part of the chain 
from a randomly chosen monomer to one of the chain ends is rotated by a random angle
about a random axis (see Ref.~\onlinecite{Linse} and references therein).
To relax the PE configuration locally, a flip
algorithm is used. In this algorithm, a randomly selected monomer 
is rotated by a random angle about the axis connecting its two neighbors
(if it is one of the end monomers, its new position is chosen randomly
at a sphere of radius $l_B$ centered at its neighbor).
The usual Metropolis algorithm is used to accept or reject the move.
For a typical value of the parameters, we run about
$10^7$ Monte Carlo steps and used the last 70\% of them to obtain statistical
averages (one Monte Carlo step is defined as the number of elementary moves
such that, on average, every particle attempts to move once).
Near the phase transition to the tail state,
the number of steps is 5 times larger.
The time for one run is typically 5 hours on an Athlon 1 Ghz computer.
Assembler language is used to speed up the calculation time 
inside the inner loop of the program.
Our code was checked by comparing with the results of 
Ref.~\onlinecite{Linse} and Ref.~\onlinecite{Pincus} 
and some references therein.

Two different initial conformations of the PE are used to make sure that
the system is in equilibrium. In the first initial conformation, 
the PE forms an equidistant coil around the macroion.
In the second initial conformation, the PE makes a straight rod.
Both initial conformations, within statistical uncertainty, give
the same values for all the calculated properties of the systems such
as the total energy, the end-end distance of the PE, the number of
collapsed monomers and the critical length $L_c$. 

An important aspect of the simulation is to determine the length of the
tail and the amount of monomers residing at the macroion surface.
In the literature, one usually defines a monomer as collapsed on the surface
if it is
found within a certain distance from it. 
This distance is arbitrarily chosen to be about two or three PE bond lengths. 
In the Appendix, we suggest an alternative more systematic method 
of determining the number of collapsed monomers.

Let us now describe the results of our Monte-Carlo simulations.
We study the collapsed length $L_1$ as a function of $L$ for
the case the macroion has radius $R=4l_B$ and charge $Q=100e$. 
This corresponds to ${\cal L}/R = 25$, exactly the same value as the one
used in Fig. \ref{pred}. The result of our simulation
is presented in Fig. \ref{lfig} together with the theoretical curve of 
Fig. \ref{pred}.
The phase transition is observed at the chain length of 142 monomers and
the critical tail length is about 16 monomers, which agrees very well with
our predictions $L_c=142$ and $L_{2c}=16$. 

We also study the case of a  salty in solution. 
As everywhere in this paper, we assume that screening by monovalent
salt can be described in the linear Debye-H\"{u}ckel
approximation. Therefore, in our simulation,
we replace the Coulomb potential of the macroion $Q/Dr$
by the screened potential
\beq
V(r)=\frac{Qe^{R/r_s}}{1+ R/r_s} \frac{e^{-r/r_s}}{Dr}~~,
\label{potential}
\eeq
where $r_s$ is the 
linear Debye-H\"{u}ckel screening length. All PE monomers are still considered
as point-like charges and Yukawa potential, $r^{-1}e^{-r/r_s}$, is used
to describe their interaction. The result of our simulation 
for the case $r_s=5l_B$
is plotted by the solid square in Fig. \ref{lfig}.
As predicted above, screening increases the maximum charge inversion 
ratio to 63\%.

Simulation at $r_s=4l_B$ shows even bigger charge inversion with
70\% ratio. This suggests that the maximum in charge inversion is located
at even smaller screening radius. However, we did not try to run the 
simulation at smaller $r_s$ in order to find
the maximum in charge inversion because, at smaller $r_s$,
the identification of adsorbed monomers becomes less unambiguous.
%

The better-than-expected agreement between MC results and theoretical
prediction of the critical length $L_c$ for the $r_s=\infty$ case
is somewhat accidental because
in Fig. \ref{lfig} we compared a zero temperature theory with a
finite temperature Monte-Carlo simulation. The temperature affects $L_c$ 
because tail monomers have 
smaller entropy compared to collapsed monomers. The self repulsion
of the tail and the repulsion from the overcharged sphere
limits the configuration space of the tail monomers, while at the
macroion, the PE self-energy is screened at the distance $A$, so
that the collapsed monomers have larger configuration space.
Therefore, the free energy is
gained when more monomers collapse on the macroion surface.
This helps to push the critical length $L_c$ to a higher value
than its value at zero temperature.

For clarifying the  role of temperature,
we carry out simulations at different $T$ and extrapolate
$L_c$ to $T=0^{\,\rm{o}}{\rm K}$. The results are shown in Fig. \ref{lctfig}.
The extrapolated $L_c$ is 134 which is 6\% lower than the 
zero temperature theoretical
prediction of 142. Also from this figure, one can see that the temperature
dependence of $L_c$ is linear. A simple analysis of the Monte-Carlo data
shows that
the entropy per monomers gained at the surface is about 2 at the
critical point.

On Fig. \ref{snap} we show
two typical snapshots of the system, one for the case $L=141$
(before the phase transition) and the other for $L=143$ 
(after the phase transition).
They again confirm that the tail appears abruptly near $L=L_c=142$.
One can clearly see an important aspect of the correlation effects:
PE segments of different turns stay away from each other
and locally, they resemble a
one dimensional Wigner crystal, which helps to lower 
the energy of the system.
Globally, however, the PE conformation resembles that of a tennis ball 
instead of a solenoid.
This obvious difference between observed conformation and
the theoretical solenoid-like ground
state is also related to thermal fluctuations. Solenoid structure 
is subjected to low energy long range bending modes, energy of which
is proportional to $k^4$, where $k$ is the wave vector of such mode. 
It is easy to show that at the room temperature with our parameters
of the system,
modes with $k\sim R^{-1}$ are strongly excited and they ``melt" the solenoid.
However, modes with large $k$ are not excited and, therefore,
the short range order between PE turns is preserved. This leads to
a compromised ``tennis ball" conformation instead of a solenoid. 
The difference in energy between a ``tennis ball" and a solenoid
conformation, however, is small compare to the interaction
between the sphere and the PE. This helps to explain the small
difference between the results of the finite temperature 
Monte-Carlo simulation and our zero temperature theory.


Monte-Carlo results similar to  Fig. \ref{lfig} for unscreened case were
independently obtained in Ref.~\onlinecite{Stoll}. 
For the screened case, however, the authors of Ref. \onlinecite{Stoll}
claimed that charge inversion reaches maximum when $r_s\simeq 3R$ 
which is still very large,
much larger than what is observed in our simulations.
This is because instead
of the Overbeck potential (\ref{potential}), the
authors of Ref. \onlinecite{Stoll} use
the Yukawa potential $Qr^{-1}e^{-r/r_s}$
for the macroion, where $r$ is the distance to the center of the macroion.
This means that they put the net charge of the macroion at the center
and screen it {\rm inside} the macroion body. As a result, 
the apparent surface charge of the macroion becomes very small and
charge inversion disappears.  New simulations\cite{Pierre} 
carried out by the same authors
using the proper potential (\ref{potential}) 
are in agreement with our theory and Monte Carlo simulations.


Before concluding this paper, we would like to mention that in our
simulation, counterion condensation on the sphere with inverted charge 
was neglected. As stated in the end of Sec. II, this is valid
if 
$\eta \ll \eta_c$.
In our Monte-Carlo simulations $\eta$ is equal to $\eta_c$
therefore, in order to study the effect of screening on charge inversion,
we choose to simulate the system 
at small $r_s \sim A$ where condensation is not very important.

In conclusion, we have studied charge inversion
for the complexation of a PE
with a spherical macroion. We started from description 
of the correlated ground state configuration 
of PE at the macroion surface instead of 
smearing of the PE charge at the surface.
As a result, we have eliminated the unphysical finite
charge adsorption
at the neutral sphere. 
Our Monte-Carlo simulations 
confirm that correlations are the driving force of 
charge inversion.

\acknowledgments
The authors are grateful to
A. Yu. Grosberg for many useful discussions and to S. Stoll
and P. Chodanowski for the possibility to read their paper
\onlinecite{Stoll} before the publication.
This work was supported by NSF DMR-9985985.

\appendix
\section{The number of collapsed monomers of polyelectrolyte.}
To better determine the number of collapsed monomers in Monte-Carlo simulation,
we use the following procedure.

Firstly, we draw the histogram of the number of
monomers found within a distance $r$ from the macroion surface
during a simulation run.
Up to a normalizing factor, this histogram is nothing but the probability
$P_r(n)$ of finding
$n$ monomers within a distance $r$ from the sphere surface.
%
%
Secondly, at a given $r$, 
we define the value of $n$ corresponding to 
the maximum in this histogram as the most likely number of monomers 
$n(r)$ found inside the distance $r$ from
the macroion surface.

Now, we show that much can be learned by 
plotting $n(r)$ as a function of $r$. 
In Fig. \ref{nofr}a, $n(r)$ is plotted for two typical cases of 
$L=140$ (before the phase transition)
and $L=150$ (after the phase transition).

Clearly, as one goes away from the macroion surface,
or $r$ grows, at first one see a rapid increase in the
number of monomers $n(r)$ found. After a distance of about two bond lengths,
this increase is slowed and stopped. It is easy to identify
the first range of $r$, where one observes a rapid increase in $n(r)$
as the collapsed layer. For the case of $L=140$, as $r$ increases
beyond this layer, $n(r)$ is always equal
to the total number of monomers $N=140$. This is the indication of a collapsed
state where all PE monomers lie in the collapsed layer near the
macroion surface. The situation
is completely different in the case of $L=150$ where beyond the
collapsed layer one sees a linear increase in $n(r)$
until $r=19$ (not shown) where $n(r)$ saturates at the maximum possible value
of 150. This is an indication of a tailed state. The slope of
the increase in this second range also provides a valuable information
on the conformation of the tailed state. As one can see, this slope is
very close to unity, what clearly indicates a one-tail state. 
This is in
agreement with our prediction that after the phase transition
the complex is in one tail state and in disagreement with the conclusion of 
Ref. \onlinecite{Pincus} that the system fluctuates between one tail
and two tail conformations (for a two tail state the slope would be 2).

A closer look at the tail part of Fig. \ref{nofr} shows that 
the slope of the tail part of $n(r)$ actually is slightly larger
than unity and grows with $r$.
This could be expected. The PE tail near the overcharged macroion is
strongly stressed in the electric field of the macroion's inverted
charge. Farther from the macroion, this electric field is weaker
and due to the thermal motion of the monomers, more than one monomer can be
found as $r$ increases by one bond length.

The final step in determining the number of collapsed monomers 
in the tailed state is accomplished by fitting the tail part of 
$n(r)$ by an empirical quadratic equation $ax^2+bx+c$.
The intersection of this curve with the $y$ axis gives the
number of collapsed monomers or the collapsed length $L_1$. 
For e.g., at $L=150$, the value of $a$, $b$ and $c$ are
0.01, 1.22 and 125 respectively (see Fig. \ref{nofr}b), 
so that the slope at the macroion surface $x=0$ is $1.22$ and the
amount of collapsed monomers is 125. 
Also, as $L$ increase, the tail gets longer and becomes more
stressed due to its self-energy, the slope of $n(r)$ decreases and
is closer to 1. The fitted value for $b$ are 1.32 at $L=145$, 1.22 at
$L=150$ and 1.09 at $L=165$.

Near the phase transition point $L=142$
one sees two maxima in the histogram $P_r(n)$ instead of one. 
One of these maxima
behaves exactly as that of one tail configuration (linearly increases
with $r$ after the collapsed layer). The other maximum behaves exactly
as that of the collapsed state (constant and equal the total
number of monomers $N=142$ after the collapsed layer). 
This is because near the phase transition the
PE fluctuates between the collapsed and the tailed state. 

\newpage

\begin{figure}
\caption{The PE winds around a spherical macroion.
Due to their Coulomb repulsion, neighboring turns lie parallel to each other.
Locally, they resemble a one-dimensional 
Wigner crystal with the lattice
constant $A$.}
\end{figure}
\begin{figure}
\caption{Schematic plots of the free energy
as function of the collapsed length $L_1$ at
different values of $L$: a) ${\cal L}<L<L^*$, b) $L^*<L<L_c$,
c) $L>L_c$.}
\end{figure}
\begin{figure}
\caption{
The collapsed length $L_1$
(solid line) and the tail length $L_2$ (dashed line)
{\it vs.} the total PE length $L$.
A first order phase transition happens at
$L=L_c$ where a tail with a finite length $L_{2,c}$
appears.} 
\label{pred}
\end{figure}
\begin{figure}
\caption{The first order phase transition to the tailed state with
increasing $L$ at ${\cal L}/R=25$. The solid line is
the theoretical prediction of the collapsed length $L_1$
as function of the PE length $L$ (same as the one plotted in Fig. 3).
The solid circles are MC results at $r_s=\infty$. 
The solid squares are MC results at $r_s=5l_B$. The dotted line is a guide
to the eyes.}
\label{lfig}
\end{figure}
\begin{figure}
\caption{The critical length $L_c$ as a function of temperature.
Because the entropy is proportional 
to the number of collapsed monomers,
a linear fit (the dashed line) is used 
to extrapolate to zero temperature.
The line has equation $y=0.029*x+133.61$.
Thus $L_c \simeq 134$ at $T=0^{\,\rm{o}}{\rm K}$.}
\label{lctfig}
\end{figure}
\begin{figure}
\caption{Two snapshot of the system for the 
cases $L=141$ (right sphere) and
$L=143$ (left sphere). }
\label{snap}
\end{figure}
\begin{figure}
\caption{The most likely number of monomers $n(r)$ 
found within a distance $r$ (measured in units of $l_B$) from
the macroion surface. a) two typical plots of $n(r)$: 
one for the case $L=140$  (below the transition length $L_c=143$) and
the other for the case $L=150$ (above $L_c$).
b) quadratic fit for the tail part of $n(r)$ for $L=150$.
The dotted line is the fitted function $f(x)=0.0098x^2+1.22x+125$.}
\label{nofr}
\end{figure}

\newpage

\begin{center}

\large{FIGURE 1}

\vspace{3cm}
\epsfxsize=12cm \centerline{\epsfbox[0 0 330 140]{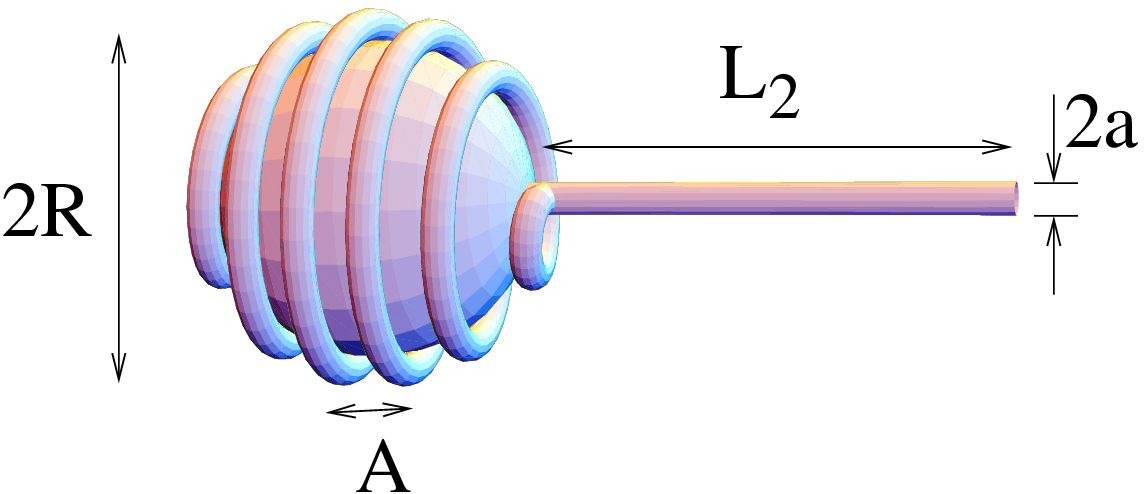}}
\newpage

\large{FIGURE 2}

\vspace{3cm}
\epsfxsize=12cm \centerline{\epsfbox{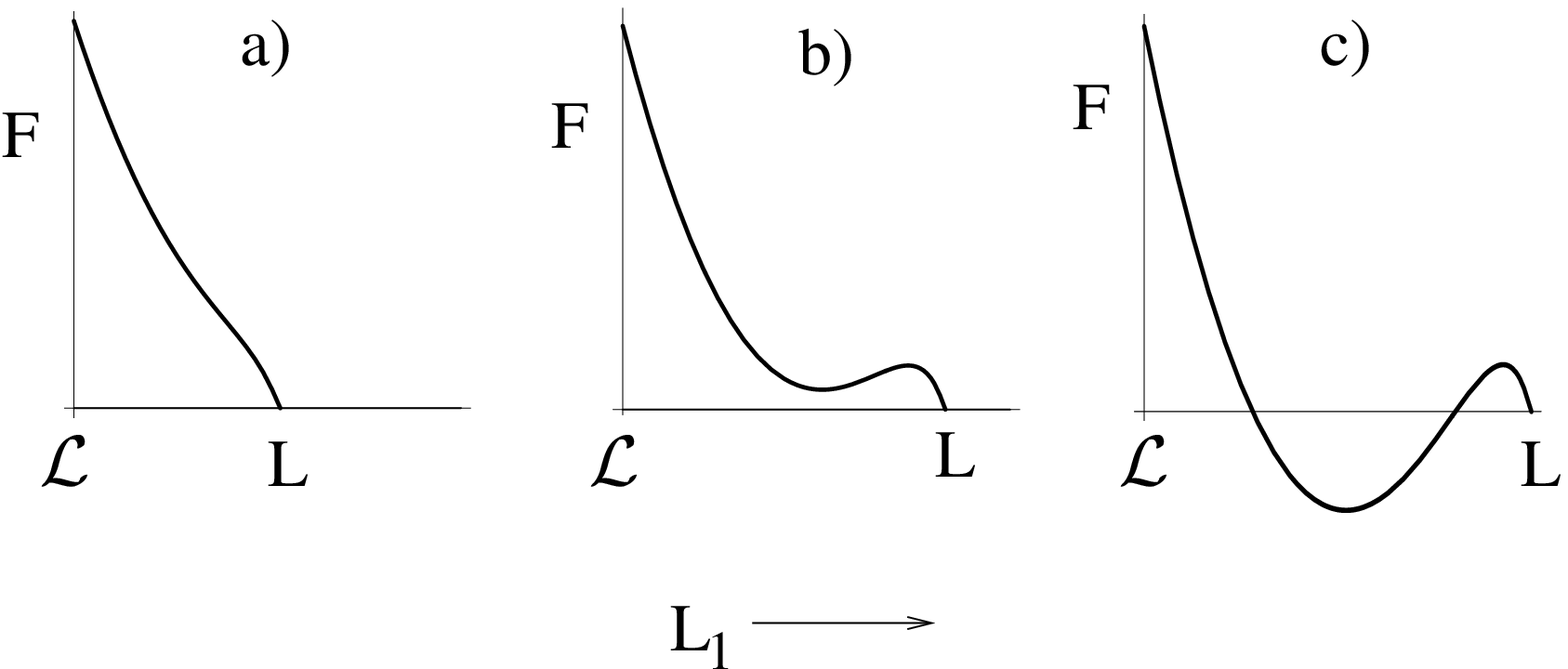}}
\newpage

\large{FIGURE 3}

\vspace{3cm}
\epsfxsize=12cm \centerline{\epsfbox{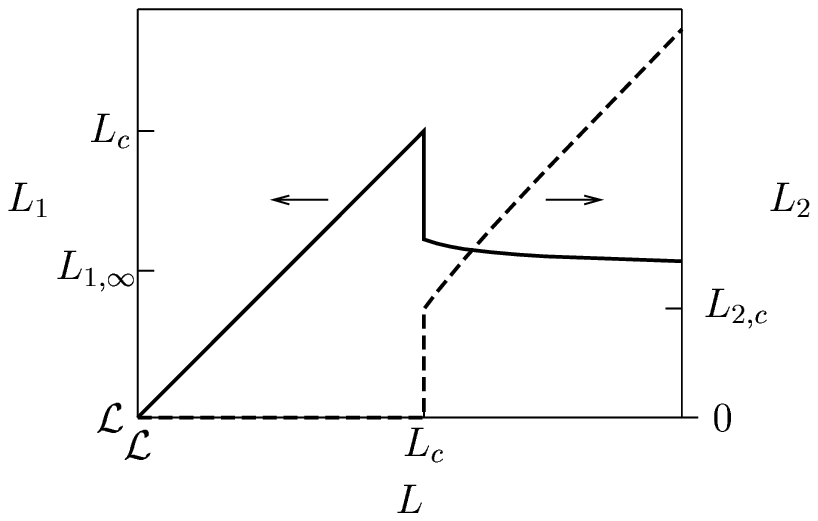}}
\newpage

\large{FIGURE 4}

\vspace{3cm}
\epsfxsize=12cm \centerline{\epsfbox{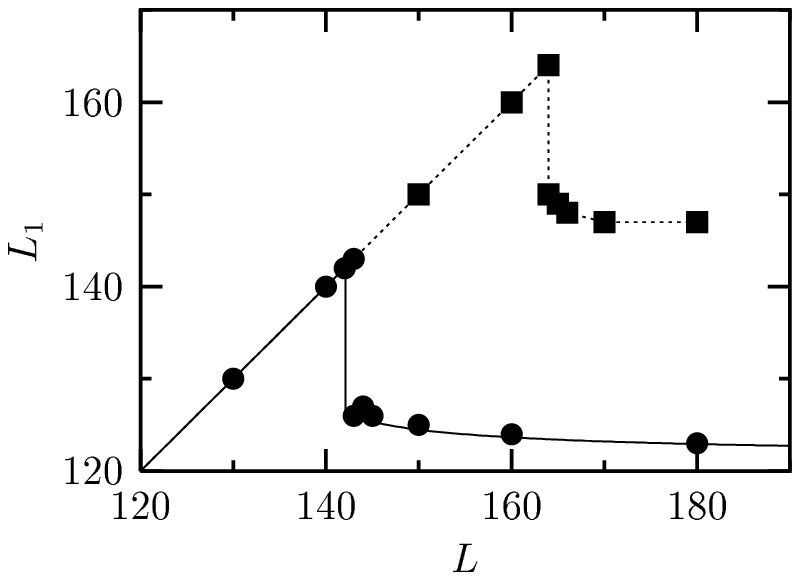}}
\newpage

\large{FIGURE 5}

\vspace{3cm}
\epsfxsize=12cm \centerline{\epsfbox{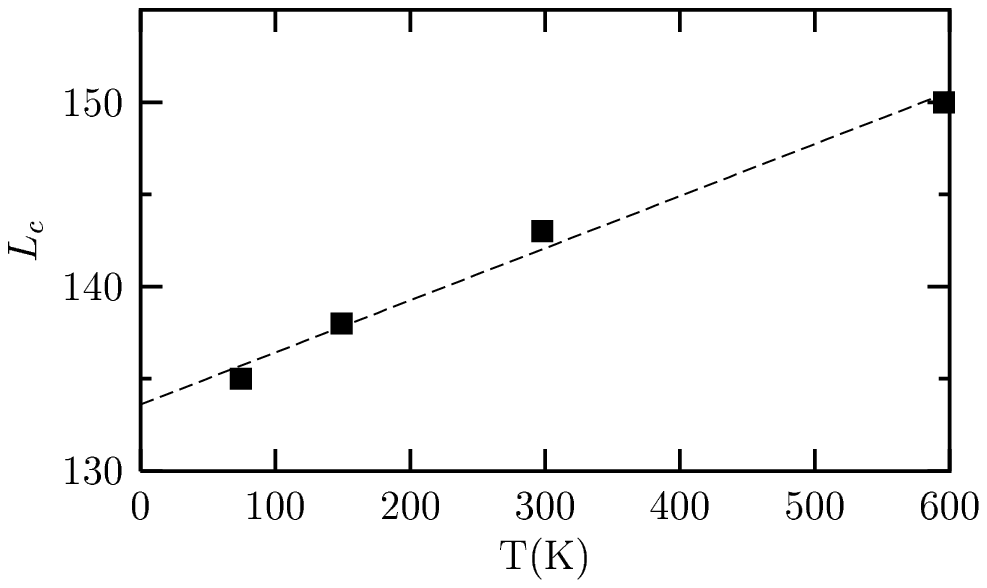}}
\newpage

\large{FIGURE 6}

\vspace{3cm}
\epsfxsize=12cm \centerline{\epsfbox{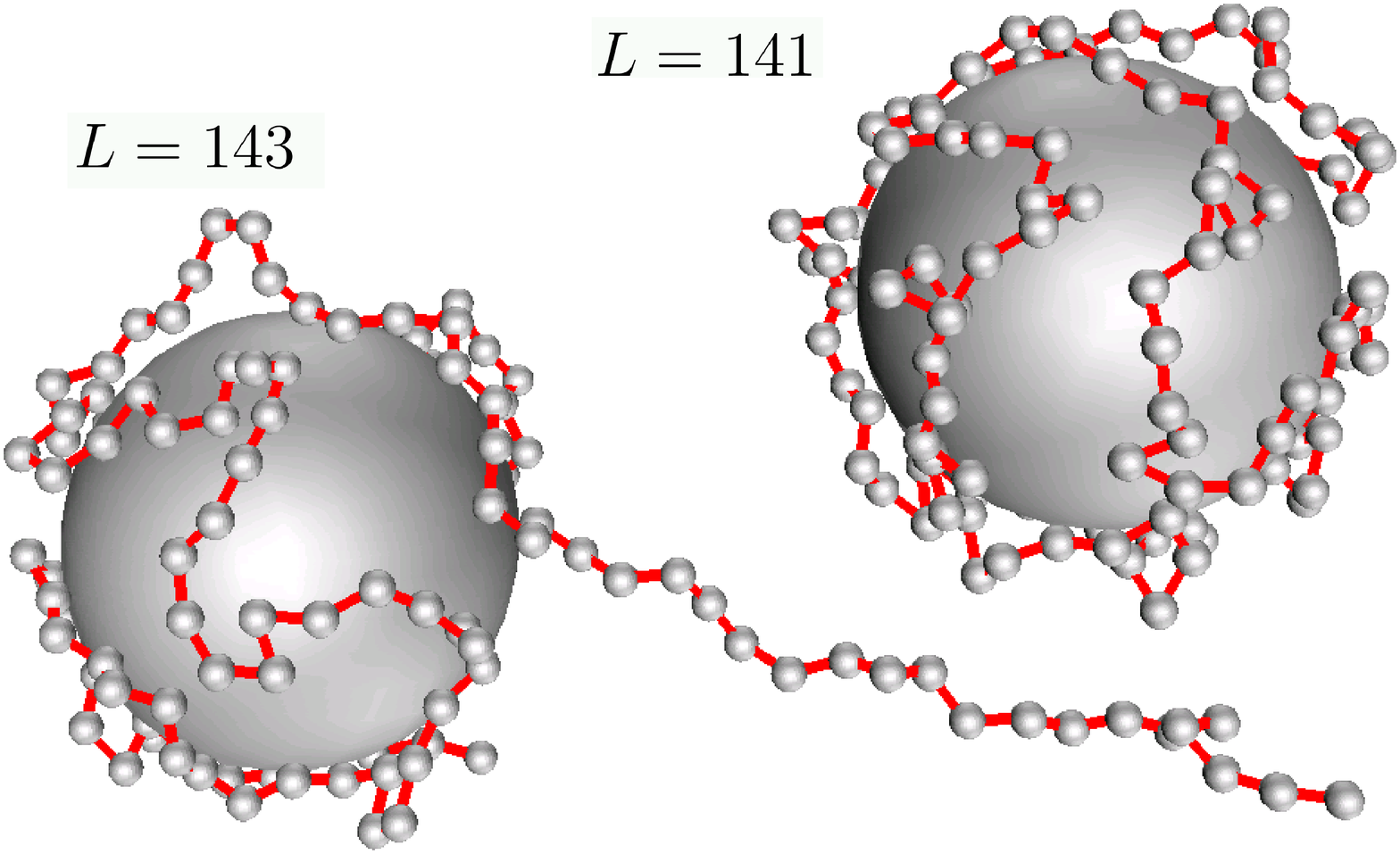}}
\newpage

\large{FIGURE 7}

\vspace{3cm}
\epsfxsize=12cm \centerline{\epsfbox{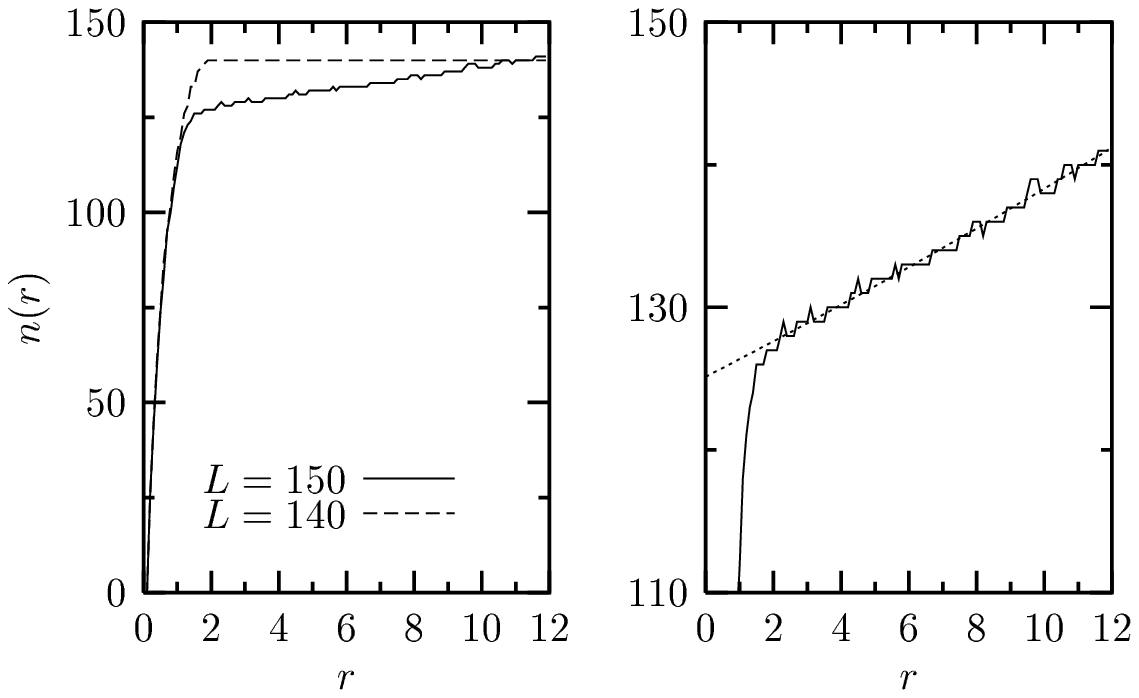}}
\end{center}


\begin{references}
\bibitem{Linse}
  Wallin T. and Linse P., 
  Langmuir {\bf 12}, (1996) 305.
\bibitem{Dubin}
  Wang Y., Kimura K., Huang Q., Dubin P. L. and Jaeger W.,
  Macromolecules {\bf 32} (1999) 7128.
\bibitem{Pincus}
  Mateescu E. M., Jeppersen C. and Pincus P.,
  Europhys. Lett. {\bf 46} (1999) 454.
\bibitem{Bruinsma}
  Park S. Y., Bruinsma R. F. and Gelbart W. M.,
  Europhys. Lett. {\bf 46} (1999) 493.
\bibitem{Joanny}
  Netz R. R. and Joanny J. F.,
  Macromolecules, {\bf 32} (1999) 9026.
\bibitem{Sens}
  Sens P. and Gurovitch E.,
  Phys. Rev. Lett., {\bf 82} (1999) 339.
\bibitem{Shklov99}
  Perel V. I. and Shklovskii B. I.,
  Physica A, {\bf 274}(1999) 446;
  Shklovskii B. I.,
  Phys. Rev. E, {\bf 60} (1999) 5802.
\bibitem{Nguyen}
  Nguyen T. T., Grosberg A. Yu. and Shklovskii B. I.,
  J. Chem. Phys.,{\bf 113}(2000) 1110.
\bibitem{Rubin}
  A. V. Dobrynin, A. Deshkovski and M. Rubinstein,
  Macromolecules 2000.
\bibitem{Gueron} M.\ Guerom, G.\ Weisbuch, Biopolimers,
  {\bf 19}, 353 (1980); S.\ Alexander, P.\ M.\ Chaikin,
  P.\ Grant, G.\ J.\ Morales, P.\ Pincus,
  and D.\ Hone, J.\ Chem.\ Phys. {\bf 80}, 5776 (1984);
  S.\ A. Safran, P.\ A.\ Pincus, M.\ E.\ Cates, F.\ C.\ MacKintosh,
  J.\ Phys. (France) {\bf 51}, 503 (1990);
  L.\ Belloni, Doctoral thesis, University of Paris IV (1982);
  Chem.\ Phys. {\bf 99} 43 (1985).
\bibitem{Stoll}
  Chodanowski P. and Stoll S.,
  Macromolecules 2000.
\bibitem{Pierre}
  Chodanowski P., PhD thesis, University of Geneva, 2001.
\end{references}
\end{document}